\documentclass[acmsmall, 10pt]{acmart}

\usepackage[normalem]{ulem}

\usepackage{tikz}
\usepackage{textcomp} 

\usepackage{multirow}
\usepackage{makecell}
\usepackage{verbatim}
\usepackage{paralist}
\usepackage[labelformat=simple]{subcaption}
\usepackage{graphicx}
\usepackage{balance}
\usepackage{rotating}
\usepackage{pifont}
\usepackage{tikz}
\usepackage{color}
\usepackage{xcolor}
\usepackage{framed}
\usepackage[para,flushleft,online]{threeparttable}
\usepackage{paralist}
\usepackage{flushend}
\usepackage{listings}
\usepackage{colortbl}
\usepackage{tabulary}
\usepackage{comment}
\usepackage{xspace}

\usepackage{wasysym}

\usepackage{
  color,
  float,
  epsfig,
  wrapfig,
  graphics,
  graphicx,
  subcaption
}

\usepackage{eucal}

\usepackage{booktabs}

\usepackage{url}

\definecolor{Gray}{gray}{0.9}

\newcommand{\jun}[1]{{\footnotesize\color{cyan}[Jun: #1]}}

\definecolor{dkgreen}{rgb}{0,0.6,0}
\definecolor{gray}{rgb}{0.5,0.5,0.5}
\definecolor{mauve}{rgb}{0.58,0,0.82}

\newcommand{\eg}{e.g.,\xspace}
\newcommand{\etal}{et al.\xspace}

\newcommand{\code}[1]{{\fontfamily{lmtt}\selectfont{#1}}}

\definecolor{cadmiumgreen}{rgb}{0.0, 0.42, 0.24}

\makeatletter
\def\thickhline{%
  \noalign{\ifnum0=`}\fi\hrule \@height \thickarrayrulewidth \futurelet
   \reserved@a\@xthickhline}
\def\@xthickhline{\ifx\reserved@a\thickhline
               \vskip\doublerulesep
               \vskip-\thickarrayrulewidth
             \fi
      \ifnum0=`{\fi}}
\makeatother

\newlength{\thickarrayrulewidth}
\usepackage{tabularx}
\usepackage{graphicx}
\usepackage{float}
\usepackage{rotating}
\usepackage{comment}
\usepackage{algorithmicx}
\usepackage{soul}
\usepackage{algpseudocode}
\usepackage{makecell}
\usepackage{enumitem}

\usepackage{tikz}

\definecolor{mygreen}{rgb}{0,0.6,0}
\definecolor{mygray}{rgb}{0.5,0.5,0.5}
\definecolor{mymauve}{rgb}{0.58,0,0.82}

\usepackage[ruled]{algorithm2e} 

\SetAlFnt{\small}
\SetAlCapFnt{\small}
\SetAlCapNameFnt{\small}
\SetAlCapHSkip{0pt}
\IncMargin{-\parindent}

\acmJournal{POMACS}
\authorsaddresses{}

\author{Jinghan Sun}
\affiliation{%
  \institution{University of Illinois Urbana-Champaign}
  \country{USA}}
\email{js39@illinois.edu}

\author{Shaobo Li}
\affiliation{%
  \institution{University of Illinois Urbana-Champaign}
  \country{USA}}
\email{shaobol2@illinois.edu}

\author{Jun Xu}
\affiliation{%
  \institution{University of Utah}
  \country{USA}}
\email{junxzm@cs.utah.edu}

\author{Jian Huang}
\affiliation{%
  \institution{University of Illinois Urbana-Champaign}
  \country{USA}}
\email{jianh@illinois.edu}

\renewcommand{\shortauthors}{Jinghan Sun, Shaobo Li, Jun Xu, \& Jian Huang}

\begin{document}

\title{The Security War in File Systems: An Empirical Study from A Vulnerability-Centric Perspective}

\date{}
\begin{abstract} 
This paper presents a systematic study on the security of modern file systems, 
following a vulnerability-centric perspective. Specifically, 
we collected 377 file system vulnerabilities committed to the CVE database in the past 20 years. 
We characterize them from four dimensions that include why the vulnerabilities appear, 
how the vulnerabilities can be exploited, what consequences can arise,
and how the vulnerabilities are fixed. This way, we build a deep understanding of 
the attack surfaces faced by file systems, the threats imposed by the attack surfaces, 
and the good and bad practices in mitigating the attacks in file systems. We envision that our study 
will bring insights towards the future development of file systems, the enhancement of file system security, 
and the relevant vulnerability mitigating solutions.


\end{abstract}

\maketitle

\thispagestyle{empty}

\section{Introduction}
\label{sec:intro}

After decades of development since the 1960s, 
file systems have become a core component in nearly
all computer systems. Working as the direct interfaces
to user data and typically running at a high privilege, 
file systems are highly security-sensitive. This has 
incentivized the creation of various techniques 
to harden the security of file systems, ranging from 
formal verification~\cite{yang-osdi04, sigurbjarnarson-osdi16, chen-sosp15, chen-sosp17} to access control
~\cite{smalley2001implementing, accesscontrol} and sanity
checking~\cite{gunawi2007improving, kim:sosp2019, min-sosp15}. 

Despite the above efforts, the security of file systems
is still frequently compromised ~\cite{ahmad2018obliviate,
Exploiti2:online}, causing damages such as data breaches 
and hijacked execution. 
To gain insights towards changing this situation, 
it is beneficial to build a systematic understanding 
of the security of file systems, from angles such 
as the attack surfaces in file systems and the threats 
imposed by the attack surfaces. Past studies
~\cite{lu-fast13,Pillai-osdi14} on file systems mostly 
focus on functionality-related perspectives, 
failing to provide the desired security-centric understanding.
Prior study~\cite{anonymous, lu-fast13} took an initial effort to 
investigate the filesystem bugs and security in Linux. 
However, it targeted solely 
on generic file systems, with a focus on the causes and consequences, 
which lacks the analysis of the whole lifecycle of the vulnerability exploitation 
and fixes processes. And we believe a thorough understanding of the attack 
exploitation procedure in file systems could help developers to advance 
the design principles and implementation strategies of file systems. 

In this paper, we present an in-depth study on the 
security of various file systems, 
aiming to fundamentally understand the entire vulnerability exploitation procedure and 
possible fix strategies. 
Specifically, our study centers around 
addressing the following questions:
\begin{itemize}[leftmargin=*]
    \item 
    \textit{Q1: What are the major attack surfaces in file systems?}
    \item 
    \textit{Q2: What are the threats posed by the attack surfaces?}
    \item 
    \textit{Q3: What are the common approaches to exploiting these attack surfaces by attackers?}
    \item 
    \textit{Q4: How are the attack surfaces fixed?}
\end{itemize}


\noindent{\bf Methodology:} To answer the above 
questions, our study takes a vulnerability-centric 
approach. We collected 377 file system vulnerabilities 
committed to the CVE database~\cite{cve} in the past 
20 years. They reported explicit 
security implications and covered nearly all the 
mainstream file systems (\emph{generic file systems} 
like Ext4~\cite{ext4}, XFS~\cite{xfs}, and F2FS
~\cite{f2fs}, \emph{mobile file systems} 
like HFS~\cite{hfs} and 
APFS~\cite{apfs}, and \emph{networked file systems} 
like NFS~\cite{nfs}, GlusterFS~\cite{glusterfs}, 
and Ceph~\cite{ceph}). 

We manually analyzed the 377 vulnerabilities 
and sought answers to questions \textit{Q1} - 
\textit{Q4}. First, we investigated the root 
causes of the vulnerabilities by analyzing
the vulnerable code and the corresponding patch(es). 
We then categorized the vulnerabilities based on
their root causes. This enabled us to gain 
an understanding of the attack surfaces 
opened by the filesystem vulnerabilities and 
the characteristics of the attack surfaces
(\textit{answering Q1}). Second, we manually and statically
reasoned the possible execution paths to exploit the 
vulnerabilities. In this process, we summarized the 
potential threats that adversaries can bring
by following those exploiting paths (\textit{answering Q2 and Q3}). 
Finally, we surveyed the patches to the vulnerabilities, 
unveiling the good and bad practices in mitigating the 
attack surfaces (\textit{answering Q4}).
Throughout the study, we derived 
a set of findings. 
We summarise the major ones that help answer \textit{Q1-Q4} as follows.


\noindent{\bf Attack surfaces:}
Vulnerabilities in file systems are diverse in types. Although 27\% of 
the vulnerabilities are of general kinds that may appear in 
other software systems (\eg memory errors and race conditions),  
a unique cause of these general vulnerabilities is the addition of 
new features, such as crash-consistency models into file systems.
The remaining vulnerabilities are unique to file systems
but attributed to different reasons:  

\begin{itemize}[leftmargin=*]
\vspace{-0.2ex}
\item
(1) 29\% are rooted in sanity checks that are used
to validate the states of various file system properties
(\eg namespace, inode attributes, and file category). The sanity
checks are highly semantic related, 
whose implementation is complicated and often leads to vulnerabilities.
\item
(2) 15\% are due to the lack of permission checks or
incorrect access control. Similar to sanity checks, 
permission checks and access control are semantic dependent
and highly complex. Their buggy implementations largely turn into permission and access-related 
vulnerabilities. 
\item
(3) 18\% are introduced when the functionality
is extended to support the use environments of specific file systems. 
For instance, many vulnerabilities appear in 
encryption mechanisms adopted by mobile file systems to 
defend against physical attacks.
\vspace{-0.2ex}
\end{itemize}

\noindent{\bf Threats and their exploitation procedure:}
File systems vulnerabilities are dangerous. 
As we will illustrate in Figure~\ref{fig:kernel-scores}, file system 
vulnerabilities pose a similar security threat level to that of
vulnerabilities in other OS components. 
The specific consequences after exploiting file system vulnerabilities
vary across vulnerability types. 

\begin{itemize}[leftmargin=*]
\item
(1) Exploiting general 
types of vulnerabilities (\eg memory errors) in file systems 
can largely cause their common sequences, such as denial-of-service
(DoS) and hijacked execution. An interesting observation is 
that general vulnerabilities in file systems can more often lead to
data leakage because file systems directly deal with user data. 
\item
(2) Exploiting erroneous sanity 
checks can also lead to DoS but more often data leakage, which allows attackers to initiate further attacks such as 
privilege escalation and permission bypass. 
\item
(3) Exploiting
lack of permission checks or incorrect access control produces more
predictable consequences, often including permission bypass and privilege
escalation as well as subsequent damages, such as data leakage
and data corruption. 
\item
(4) Exploiting the remaining 
vulnerabilities typically causes consequences pertaining 
to the use contexts of the file systems. For instance, 
exploiting vulnerabilities in the encryption modules 
of mobile file systems often enables physical attacks,
while exploiting vulnerabilities in the networking management 
of network file systems usually leads to code injection attacks.
\end{itemize}

\noindent{\bf Fixes with patches:} File system patches
are diverse in both complexity and functionality. 56\% of
the patches only change less than 10 lines of code, using 
simple operations such as adding a boundary check and 
fixing a syntax error (see the details in Figure~\ref{fig:patch_bar}). 
In contrast, around $20\%$ of the patches bring modifications to over 50 lines of code,
involving complex behaviors ranging from full validation 
of data structures to fixes in multiple components.
The majority of existing and current patches work in an ad-hoc style. 
For instance, the patches to many race conditions simply make the responsible 
operations atomic. These ad-hoc patches can heal the
target vulnerabilities, but they fail to prevent similar 
vulnerabilities from emerging or fundamentally mitigate them.
Nevertheless, our study identifies positive signs. 
Strategies to enforce preventive and secure-by-design patching
are arising. For instance, many patches of race conditions switch 
from lock-based implementations to lock-free implementations,
fundamentally eliminating issues like deadlocks. 
File systems are also often correlated, so as their patches. 
In particular, a patch to the VFS subsystem often requires 
updating the underlying file systems. Otherwise, the patch
may not come into effect. However, today's practice often 
overlooks the correlated file systems when patching the target ones,
thus leading to insecure patches.



\noindent{\bf Contributions:} Our main contributions are as follows.
\begin{itemize}[leftmargin=*]
    \item We perform the first in-depth study on the security of 
    file systems, from a vulnerability-centric perspective. The 
    study unveils the attack surfaces of file systems, followed
    by demystifying the root causes and threats of the attack surfaces. 
    
    \item We present 21 previously-unknown findings throughout 
    the study. Besides facilitating a deeper understanding of the attack 
    surfaces, the findings also reveal how the development of 
    file systems influences or even impedes security. 
    
    \item We distill 7 insights from the results and findings 
    of the study. The insights point out the deficiencies in today's 
    practice of addressing security threats faced by file systems 
    and also bring up a set of guidance towards improvements.
    
\end{itemize}

\section{Study Methodology}
\label{sec:meth}

\begin{table}[t]
    \footnotesize
    \caption{File systems covered in our study.} 
    \label{tab:fstype}
    \vspace{-2ex}
    \centering
    \begin{tabular}{p{60pt}<{\centering}|p{215pt}<{\centering}|p{36pt}<{\centering}}
    \hline
    \textbf{Type} & \textbf{File Systems} & \textbf{\# of CVE}  \\
      \hline
	Generic       & Ext4, VFS, XFS, F2FS, BtrFS, profs, Ext2, Ext3, tmpfs, ReiserFS, JFS  &  178               \\
	Mobile      & HFS/HFS+, APFS, F2FS, eCryptFS         &   57                     \\
	Networked   & NFS, OpenAFS, GlusterFS, CephFs, HDFS  &  142                \\
        \hline
        \textbf{Total}  &    --   & 377              \\
       \hline
    \end{tabular}
\end{table}


Our study aims to understand the attack surfaces in file systems,
from the perspectives summarized as \textit{Q1} - \textit{Q3}. 
To support the study, we consider public vulnerabilities
as the data source. In comparison to other applicable data 
sources such as attacks that happened in the past, vulnerabilities are 
more accessible and more abundant.

\subsection{Data Collection}

We collected file system vulnerabilities from the CVE database~\cite{cve}. 
We considered the CVE database for three reasons. First, the CVE reports
are reviewed and confirmed by trustworthy parties (major IT vendors, 
security companies, and research organizations~\cite{CVEReque73:online}) 
and thus, offer high reliability. Second, the CVE reports largely provide
important details, such as root cause information and patch information. 
Such details are helpful for us to answer \textit{Q1} - \textit{Q3}. 
Third, the CVE reports are large in number and diverse in terms of 
sources, which should properly reflect the overall situation.

Specifically, we examined the CVE reports committed in the past 
20 years (January 1999 to December 2020) and identified those 
associated with mainstream file systems. Older reports were omitted since 
they may not represent what is happening at present. We further refined the 
reports to only keep those carrying information about or giving 
references to (i) location of the vulnerable code, (ii) analysis 
of the root cause (comprehensive or brief), (iii) possible 
exploitation of the vulnerability and subsequent 
consequences (which will be verified following our 
approach in \S\ref{subsec:vulanalysis}),
and (iv) the fixes with patches.

\begin{figure}[!t]
        \centering
        \begin{minipage}[c]{\linewidth}
                \centering
                \includegraphics[width=0.85\linewidth]{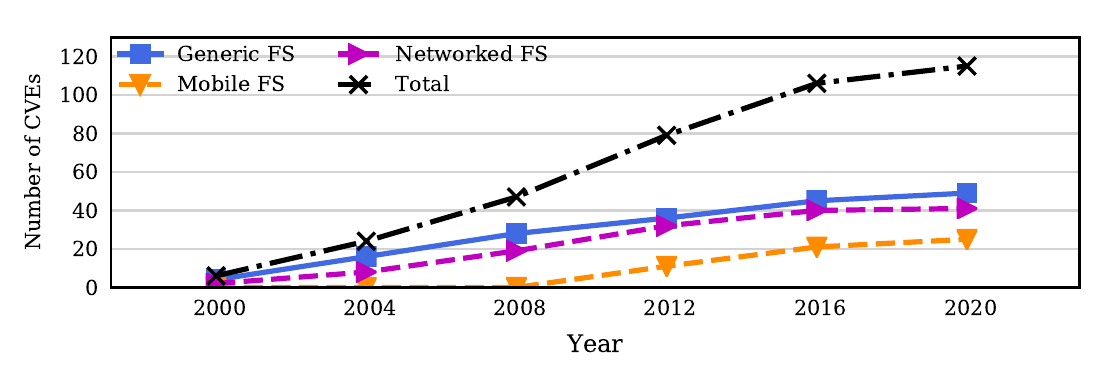}
        \end{minipage}
        \vspace{-4ex}
        \caption{Distribution of vulnerabilities over the past twenty years.}
        \label{fig:time_distribution}
\end{figure}

In the end, we collected 377 file system vulnerabilities, whose 
distribution across time is shown in Figure~\ref{fig:time_distribution}. They cover most of the widely 
deployed file systems, as illustrated in Table~\ref{tab:fstype}. The vulnerabilities also span a wide spectrum of types 
and related details are presented in \S\ref{sec:taxonomy}. 

\subsection{Vulnerability Analysis}
\label{subsec:vulanalysis}
We ran three analyses on each of the 377 vulnerabilities. 

\paragraph{\textbf{Root cause analysis}} The first analysis 
focuses on understanding the root cause of a CVE report.
In this analysis, we first extract descriptions of 
the root causes from the report and then 
verify the description through interpretation of the 
vulnerable code and its patch.


\paragraph{\textbf{Exploitation analysis}} 
Following the root cause analysis,
we aim to understand the possible exploitation of a vulnerability
and the subsequent consequences. One common approach is to reproduce
the existing proof-of-concept (PoC) exploits against the vulnerability ~\cite{mu2018understanding}. 
This approach, however, has limitations. 
First, PoC exploits are not available for many CVE reports. Omitting 
those CVE reports can downgrade the generality of our studies. 
Second, the reproduction of PoC exploits is labor intensive, 
which cannot scale to support large-scale studies like ours.

In our study, we use a more generic and more 
lightweight approach. We first parse the description
in a CVE report to retrieve the reported exploitation
and its consequences. We then reason the vulnerable code 
to identify execution paths that can verify the description.
Specifically, an execution path successfully verifying the 
exploitation must meet the following conditions:

\begin{itemize}[leftmargin=*]
    \item The execution path covers a feasible route from 
    a user-space interface to the vulnerable code.
    \item The execution path covers a feasible route to 
    triggering the vulnerability.
    \item The execution path covers a feasible route 
    from triggering the vulnerability to causing the described
    consequence(s).
\end{itemize}

\noindent If no such execution paths can be identified for a CVE report, we skip the report. The example in Figure~\ref{fig:CVE-2010-3078} illustrates our approach. 
In the example, a local data structure \emph{fa} (declared at line 2) is not fully initialized before it is copied to user space at line 7. The CVE report says 
\emph{``It allows local users to obtain potentially sensitive information from kernel stack memory via an ioctl call''.} 
 
\noindent We find that any execution path first opening a legitimate file in an XFS file system and then calling ioctl (with XFS\_IOC\_FSGETXATTR) on the 
opened file will trigger the vulnerability, and leak data from the kernel stack. This way the exploitation and the consequence are considered verified.

\lstdefinestyle{base}{
  language=C,
  emptylines=1,
  breaklines=true,
  basicstyle=\ttfamily\color{black}\small,
  moredelim=**[is][\color{cadmiumgreen}]{@}{@},
  numberstyle=\tiny,
  xleftmargin=2.5mm
}

\begin{figure}[!t]
  \centering
\begin{lstlisting}[language=C,style=base]
int xfs_ioc_fsgetxattr(xfs_inode_t *ip,int attr,...){
  struct fsxattr fa; 
@+ memset(&fa, 0, sizeof(struct fsxattr));@
  // data structure ``fa'' is uninitialized before patching
  xfs_ilock(ip, XFS_ILOCK_SHARED);
  fa.fsx_xflags = xfs_ip2xflags(ip);
  if (copy_to_user(arg, &fa, sizeof(fa)))
    return -EFAULT; 
}
\end{lstlisting}
\vspace{-2ex}
  \caption{An uninitialized memory vulnerability in XFS and its patch (CVE-2010-3078).}
  \label{fig:CVE-2010-3078}
        \vspace{-2ex}
\end{figure}


\paragraph{\textbf{Patch analysis}} The last analysis examines 
the patches to file system vulnerabilities. One goal is to 
understand the strategies used by the existing patches (\eg ad-hoc or systematic). 
This way, we expect to unveil whether today's practice of file system 
patching  can help fundamentally resolve the issues of vulnerabilities.
Another goal of the analysis is to summarise the aspects
overlooked by file system developers when patching vulnerabilities.
We anticipate bringing related evidence to incentivize broader attention to those aspects.

\subsection{Threats to Validity}
While we carefully designed our methodology, it may still pose threats to the validity of our findings. First, similar to other sampling-based studies, our study can miss representative vulnerabilities or even miss certain types of vulnerabilities. Thus, our findings might be biased towards the collected samples. We admit the possibility of this threat but we should have reduced it by adopting the most comprehensive, most standard dataset. Second, we verify the exploitation of a vulnerability and the consequences through static reasoning. Unlike PoC exploits, static reasoning does not guarantee the fidelity of the results. As such, we may present underestimated/exaggerated/erroneous threats of the vulnerabilities. To mitigate this threat, we document the execution paths identified to exploit the vulnerability and then have another person review and vet the execution paths. Any execution paths that raise discrepancy are discarded. Third, the analysis of the patches can often require deep domain knowledge. Lack of such knowledge will lead to misunderstanding about the patches. To mitigate the threat, we only allow authors whose research focus is file/storage system to look at the patch and again, we have multiple authors to vet the results.

\begin{table*}[t]
    \footnotesize
    \caption{Summary of file system vulnerabilities. \textit{M:} Memory Errors; \textit{C:} Concurrency Issues; \textit{S:} Sanity Check Errors;  \textit{P:} Permission Errors; \textit{N:} Network Errors; \textit{O:} Others (\eg hash collisions like \texttt{CVE-2014-7283}).}
    \label{tab:genericfs}
    \vspace{-2ex}
    \centering
    \resizebox{.9\textwidth}{!}{
    \begin{tabular}{m{34pt}<{\centering}|m{49pt}<{\centering}|p{25pt}<{\centering}|p{18pt}<{\centering}|p{16pt}<{\centering}|p{16pt}<{\centering}|p{16pt}<{\centering}|p{16pt}<{\centering}|p{16pt}<{\centering}|p{16pt}<{\centering}}
      \toprule
        \multicolumn{5}{c|}{\textsc{\textbf{File Systems}}}  & \multicolumn{5}{c}{\textsc{\textbf{Vulnerabilities}}} \\ \hline
              \textbf{Type} & \textbf{Name}  & \textbf{Release Time} &\textbf{\#CVEs}  & M & C & S & P & N & O  \\\hline

        \multirow{12}{*}{Generic}  & JFS \cite{jfs} & 1990 & 4  
        &  \colorbox{pink}{50\%} & 0\% & \colorbox{pink}{50\%} & 0\% & 0\% & 0\%\\
        
        &  Ext2 \cite{ext2}                   & 1993 & 7 & 
        14\% & 14\% & \colorbox{pink}{43\%} & 14\% & 0\% &14\% \\
        
        & XFS \cite{xfs}                 & 1993 & 23
        & \colorbox{pink}{30\%} & 4\% & \colorbox{pink}{35\%} & 4\% & 0\% & 26\%\\
        
        &  VFS \cite{vfs}                     & 1995 & 28 &  18\%  & 11\% & \colorbox{pink}{40\%} & 18\% & 0\% & 14\% \\
        
        &    procfs \cite{procfs}     & 1999 & 18 &  \colorbox{pink}{22\%}    & 6\%& 11\% & \colorbox{pink}{22\%} & 0\% & 39\% \\
        
         & ReiserFS  \cite{reiserfs}      & 2000 & 5 & \colorbox{pink}{40\%} & 0\% & \colorbox{pink}{40\%} & 20\% & 0\% & 0\% \\


        & Ext3 \cite{ext3}               & 2001 & 11 & 9\% & 0\% & \colorbox{pink}{36\%} & 18\% & 0\% & 36\% \\
        
         & tmpfs   \cite{tmpfs}        & 2001 & 6 & 17\%    & 0\%& 17\% & \colorbox{pink}{33\%} & 0\% & 33\% \\


        & Ext4  \cite{ext4}                 & 2008 & 45 &
        13\% & 4\% & \colorbox{pink}{49\%} & 4\% & 0\% & 24\%\\

        & Btrfs  \cite{btrfs}        & 2009 & 13 &
        15\% & 8\% & \colorbox{pink}{38\%} & \colorbox{pink}{31\%} & 0\% & 8\% \\

        & F2FS    \cite{f2fs}                 & 2012 & 18 &
        11\% & 11\% & \colorbox{pink}{56\%} & 6\% & 0\% & 17\% \\ \hline
            
            
    & \textbf{Total} & \textbf{Avg $\rightarrow$} & 178  & 19\% & 6\% & \colorbox{pink}{39\%} & 13\% & 0\% & 22\%\\
    \thickhline
    \multirow{5}{*}{Mobile}  &  HFS/HFS+ \cite{hfs} & 1985 & 8 & \colorbox{pink}{25\%} & 0\% & \colorbox{pink}{75\%} & 0\% & 0\% & 0\%\\
    
        & eCryptFS \cite{ecryptfs}        & 2006 & 16 & \colorbox{pink}{44\%} & 0\% & \colorbox{pink}{25\%} & 19\% & 0\% &13\% \\

    & F2FS   \cite{f2fs}                        & 2012     & 11      & \colorbox{pink}{36\%} & 0\% & \colorbox{pink}{45\%} & 9\% & 0\% & 9\%\\
    
    & APFS \cite{apfs}   & 2016   & 4  
    & \colorbox{pink}{25\%} & \colorbox{pink}{25\%} & \colorbox{pink}{25\%} & \colorbox{pink}{25\%} & 0\% & 0\%  \\

    & Others                &       --  & 18     & 22\% & 17\% & \colorbox{pink}{33\%} & 6\% & 0\% & 22\%  \\\hline
    
    & \textbf{Total}  & \textbf{Avg $\rightarrow$} &
    57 & \colorbox{pink}{32\%} & 7\% & \colorbox{pink}{39\%} & 11\% & 0\% & 12\%\\
    \thickhline


  \multirow{6}{*}{Networked} & NFS \cite{pawlowski2000nfs}           & 1984   & 72 & 17\% & 7\% & 13\% & 20\% & \colorbox{pink}{24\%} & 19\%  \\
    & GlusterFS \cite{glusterfs}  & 2005  &  18 & 16\% & 0\% & 11\% & 11\% & \colorbox{pink}{39\%} & 22\% \\
    
     & OpenAFS \cite{openafs} & 2006  & 27
    & \colorbox{pink}{44\%} & 4\% & 11\% & 11\% & \colorbox{pink}{22\%} & 7\% \\

    & HDFS \cite{hdfs}	   		& 2006  & 6 & 0\% & 0\% & 33\% & \colorbox{pink}{67\%} & 0\% & 0\% \\
    
    & CephFS \cite{weil2006ceph}  & 2012  & 19 & 16\% & 0\% & 16\% & 21\% & \colorbox{pink}{26\%} & 21\% \\
    
\hline
        
         & \textbf{Total}  & \textbf{Avg $\rightarrow$} & 142 & 21\% & 4\% & 13\% & 19\% & \colorbox{pink}{25\%} & 16\%\\
        \thickhline
    \end{tabular}
    \vspace{-3ex}
    } 
\end{table*}

\section{Taxonomy of File System Vulnerabilities}
\label{sec:taxonomy}

Vulnerabilities appear in every mainstream file system and their discovery also presents an increasing trend, as shown in Table~\ref{tab:genericfs} and Figure~\ref{fig:time_distribution}. 
To gain better insights, the first deeper understanding we build about 
is the categories of vulnerabilities. We categorize the vulnerabilities into two types: 

\begin{itemize}[leftmargin=*]
    \item \emph{Generic vulnerabilities} -- vulnerabilities that can appear in any kinds of software systems. The major types include memory errors and race conditions.  
    
    \item \emph{Semantic vulnerabilities} -- vulnerabilities that only appear in file systems but can appear in all kinds of file systems. These vulnerabilities are mostly caused by cursoriness or errors in semantic-dependent or functionality-dependent implementations.

\end{itemize}

The categorization brings a set of statistical findings
and interesting observations. 


\begin{figure}[!t]
        \centering
        \begin{minipage}[c]{\linewidth}
                \centering
                \includegraphics[width=0.85\linewidth]{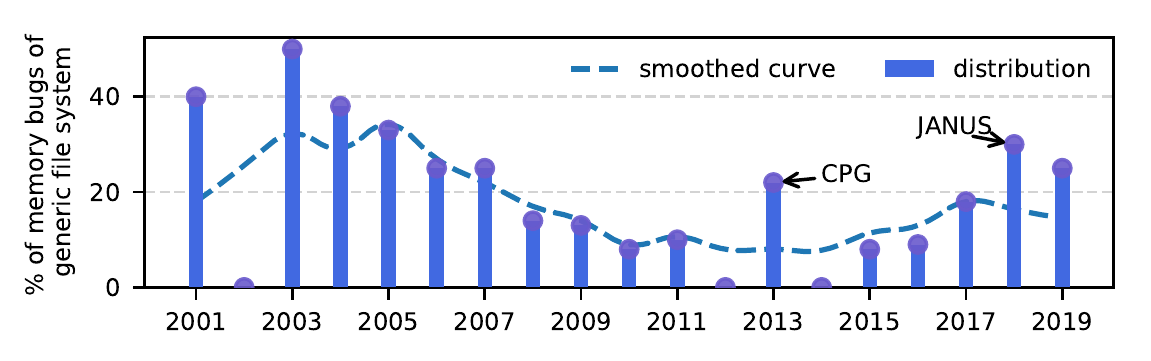}
        \end{minipage}
        \vspace{-2ex}
        \caption{Distribution of memory errors reported from file systems over time. The deployment of JANUS~\cite{Xu-sp19} and Code Property Graph (CPG)~\cite{cpg} led to spikes of reported memory bugs in 2013 and 2018.}
        \label{fig:ge-me}
\end{figure}

\noindent\textbf{\texttt{F-1}: Generic vulnerabilities are common in file systems 
($>$27\%) but present positive signs. And there has been a misconception that generic vulnerabilities 
are decreasing as file systems become mature.} Our study reveals that there would be a significant amount of 
deep memory and concurrency vulnerabilities that can be detected with advanced bug-finding tools, as shown in Figure~\ref{fig:ge-me}. 
Generic vulnerabilities account for a significant portion of the vulnerabilities in file systems. 
In particular, memory errors and concurrency issues led to 21.7\% and 5.5\% of all the vulnerabilities, respectively. 
The numbers are not surprising since file systems are largely developed in memory unsafe languages (e.g., C/C++) and widely support concurrency. 

No doubt, the above numbers give the important message that generic vulnerabilities still constitute a major attack surface in file systems. 
Apparently, it is unwise to underestimate or overlook generic
vulnerabilities at this point. However, we do observe positive 
signs. Various bug-finding tools (\eg JANUS~\cite{Xu-sp19} and Code Property Graph~\cite{cpg}) and bug-detecting tools (\eg
AddressSanitizer~\cite{serebryany2012addresssanitizer} and ThreadSanitizer~\cite{serebryany2009threadsanitizer}) 
emerged in recent years. Real-world 
deployment of these tools has been disclosing generic vulnerabilities 
in file systems. Consider Figure~\ref{fig:ge-me} as an example. 
Code Property Graph~\cite{cpg} and JANUS~\cite{Xu-sp19}, 
after deployment, discover a surge of memory 
errors in generic file systems.
JANUS applies fuzzing techniques to file system vulnerability detection by mutating file system images and manipulating file system operations 
as the input to the fuzzer. The large search space of fuzzing helps detect deep memory errors that cannot be manually tested via hand-crafted images 
(e.g., \texttt{CVE-2018-10880}). This could encourage researchers to properly apply other advanced bug detection tools to file systems. 

\begin{figure}[!t]
        \centering
        \begin{minipage}[c]{\linewidth}
                \centering
                \includegraphics[width=0.85\linewidth]{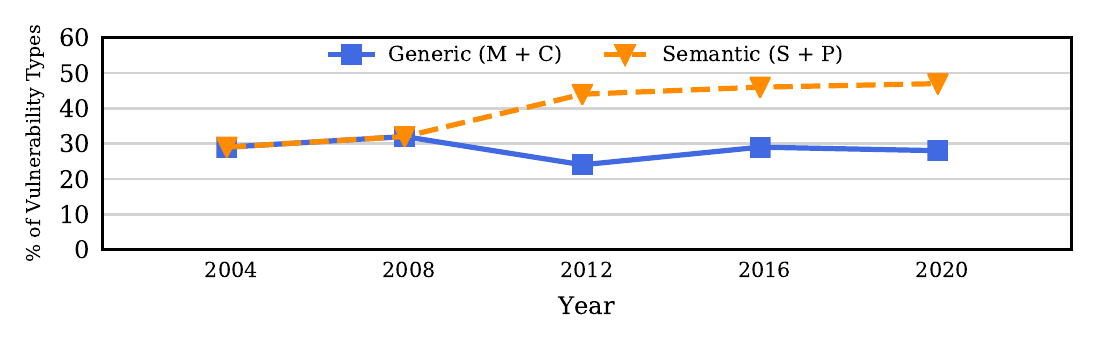}
        \end{minipage}
        \vspace{-4ex}
        \caption{Distribution of vulnerabilities across time. 
        Generic vulnerabilities are stabilizing and semantic
        vulnerabilities present an increasing trend. 
        }
        \label{fig:distribution}
\end{figure}


\vspace{0.5em}

\noindent\textbf{\texttt{F-2}: Semantic vulnerabilities are the 
dominant category of vulnerabilities in file systems (43\%).} 
File systems involve highly complex semantics for reliable and efficient file management. 
Implementation of the semantics is challenging and often 
error-prone, bringing the dominant category of vulnerabilities. 
There are two major sources leading to semantic vulnerabilities, 
sanity checks and file permissions. We separately discuss 
them in the following. 

\emph{Sanity checks.} File systems maintain a mass of 
semantic-related states, such as 
file system namespace, inode attributes, cache consistency, and so on. 
Missing sanity checks on the states is a major 
way of introducing vulnerabilities. For instance, the implementations 
of the journaling in Ext2 and Ext3 never validate the journal 
superblock before reading it, which can cause an assertion or even 
kernel corruption (\texttt{CVE-2011-4132}). Erroneous sanity checks on 
the states is another way of introducing vulnerabilities. 
For example, the f2fs utilities before version 1.12.0  
has a logical flaw when validating the sanity of the 
superblock (\texttt{CVE-2020-6070}). The logical flaw can lead 
to bypass of validation and even user-controlled execution.
In total, missing sanity checks and erroneous sanity checks
caused 29\% of the total vulnerabilities in file systems. 

\emph{File permissions.} File permissions are indispensable 
to the protection of user data. To enforce file permissions, 
file systems rely on standard permission models 
(\eg \texttt{u/g/o:r/w/x}) and access control list (ACL) 
(\eg POSIX ACLs) to manage accesses. The actual implementations 
of permission models and ACL, however, often miss or add
insufficient permission checks, leading to 15\% of the 
vulnerabilities in file systems. For example, ReiserFS misses  
permission checks on the accesses to the 
directory storing file attributes (i.e., \texttt{xattrs})
(\texttt{CVE-2010-1146}). This enables any user to access 
or modify the attributes of files that are not allowed.

In contrast to generic vulnerabilities, semantic vulnerabilities 
are more concerning today. On the one hand, most of the semantic 
vulnerabilities seem to be discovered in an ad-hoc manner. 
We did not identify many tools that systematically detect semantic vulnerabilities. 
While some generic tools, like the 
fuzzing-based JANUS, do help find semantic vulnerabilities,
they are still better at detecting generic vulnerabilities 
as they often lack domain-specific guidance to capture semantic 
vulnerabilities. On the other hand, as shown in Figure~\ref{fig:distribution},
the number of reported semantic vulnerabilities is clearly increasing 
over time. This might be attributed to that more file systems 
are included in our study as time moves forward or that 
the number of hidden semantic vulnerabilities is significantly 
large. But either way, the trend implies an increasing threat posed 
by semantic vulnerabilities.


					


\vspace{0.5em}

\noindent
\fcolorbox{black}{Gray}{\parbox[c]{13.7cm}{
\textbf{Summary:} Both generic vulnerabilities
and semantic vulnerabilities open large attack 
surfaces in file systems. However, generic vulnerabilities
preset a seemingly improving trend while semantic 
vulnerabilities show the opposite.

\vspace{0.5em}

\textbf{Insight-1:} To heal the attack surfaces of 
file systems, more efforts should be prioritized to 
mitigate semantic vulnerabilities. 
In particular,
it is wise to explore tools that can  
systematically unveil semantic vulnerabilities. 
}}

\vspace{0.5em}

Our categorization brings knowledge about the major 
vulnerabilities to be addressed. However, it gives 
few insights into how. To this end, we further 
investigate which file system functionality and 
what file operations introduce the vulnerabilities. 
In a nutshell, every major functionality of file systems
brings in vulnerabilities, but inode management, 
block management, and page cache system brings in more. 
And a key observation is that extending the existing 
core functionality with new features contributed to 
a substantial number of vulnerabilities. 
We detail our findings below. 

\vspace{0.5em}

\noindent \textbf{\texttt{F-3:} Inode management is a major source
of introducing vulnerabilities (13\% out of Linux FS).} 
As per our study, inode management is also one of the most 
vulnerable parts of file systems. One reason is the complexity 
in the nature of inode management. Another reason, 
which is becoming more dominant, is the increasing need 
to support newly emerging features such as inline data blocks and extended attributes. A particular example
is Ext4. Unlike Ext2 and Ext3 that uniformly keep extended attributes
in a block, Ext4 supports three on-disk formats to store 
extended attributes, including internal inode, additional 
block, and dedicated inode. 
In regular cases where the extended attributes are small, 
Ext4 simply stores them in the internal inode. 
However, when the attributes grow bigger, Ext4 will
swap them to an additional block. This type of format 
adjustment brings extra complexities and causes multiple 
vulnerabilities (\eg \texttt{CVE-2018-11412}). 
{Prior work proposed isolating metadata structures of unrelated files and directories to avoid failure propagation} \cite{lu-osdi14}. 
{And inode replication was also proposed to ensured its fault tolerance} \cite{inodereplication}. 
{We believe more efforts should be made to decouple the \emph{inode} management from other components in file systems.  
}




\begin{table}[!t]
    \footnotesize
    \caption{Distribution of vulnerabilities caused by different 
    features in block management. Only representative file
    systems are selected. The numbers are per file system.}
	\vspace{-2ex}
    \label{tab:fs-feature}
    \centering
    \begin{tabular}{p{30pt}<{\centering}|p{28pt}<{\centering}|p{50pt}<{\centering}|p{50pt}<{\centering}|p{50pt}<{\centering}|p{40pt}<{\centering}}
    \toprule
    \textbf{File Systems} & \textbf{Extent} & \textbf{Extended Attribute} & \textbf{Delayed Allocation} & \textbf{Flex/Meta Block Group} & \textbf{Inline Data} \\
	\hline
	Ext4 \cite{ext4} & \colorbox{pink}{29\%} & \colorbox{pink}{13\%} & 4\% & \colorbox{pink}{16\%} & 6\% \\
	XFS \cite{xfs} & 6\% & \colorbox{pink}{16\%} & - & - & -\\
	F2FS \cite{f2fs} & - & 6\% & - & - & -\\
	JFS \cite{jfs} & - & 3\% & - & - & -\\
	\thickhline
    \end{tabular}
	\vspace{-5ex}
\end{table}

\vspace{0.5em}

\noindent \textbf{\texttt{F-4:} Block management incurs plenty of
vulnerabilities (21\%) and presents an increasing trend. {These vulnerabilities could be mitigated by having a more expressive interface to offload block management to 
the storage devices.}}
Block management is another major source of introducing 
vulnerabilities. In Table~\ref{tab:fs-feature}, we present the ratio of 
vulnerabilities caused by different features in 
block management. This is not surprising since block management
has complexity comparable to inode management. Moreover, 
more features are being integrated into block management 
for better efficiency. In particular, recent file systems, 
such as Ext4 and BtrFS,  uses extents instead of
direct/indirect blocks to organize file blocks. 
The codebase of extent management (7K lines of code)
is much larger than that of direct/indirect block 
organization (1.5K lines of code).
{
The need for fast indexing, flexible block group}~\cite{flexblock} {and delayed block allocation all contribute to the increasing complexity of the block management, and hence result in more vulnerabilities.}
{Instead of purely relying on the storage software for the block management, we believe that future file systems should explore an expressive interface to offload the block management tasks 
to storage devices. This could not only ease the file system development and verification procedure, but also isolate its vulnerabilities from the OS kernel. 
As the hardware resources (i.e., storage processor and memory capacity) inside storage controllers become more powerful, the device is capable of handling more management tasks. 
The most recent work on device-level file systems, such as DevFS} \cite{devfs-fast2018} {and KEVIN} \cite{koo2021modernizing}, demonstrated the feasibility of offloading the file system 
functionalities to storage devices, however, they purely focused on performance improvement. 
It is highly desirable to explore the design trade-offs with respect to the security enhancement.

\vspace{0.5em}

\noindent \textbf{\texttt{F-5}: {Page cache causes 12\% of the vulnerabilities and they could leak sensitive kernel information.}}
Modern file systems widely use page cache to temporarily store 
data blocks that will be flushed to the persistent storage at 
certain time points. The use of page cache avoids frequent
interactions with the persistent storage and thus, improves
the performance of applications. However, the extra layer 
of cache creates new space for vulnerabilities. A major 
type of such vulnerabilities is caused by uninitialized cache. 
Specifically, developers often do not initialize 
many fields of metadata structures saved in page cache. 
Later when the page cache is flushed,
the uninitialized fields, whose memory may still carry sensitive
kernel information, will be leaked to the persistent storage 
(\eg \texttt{CVE-2005-0400 and CVE-2006-6054}). Any user who has access
to the block can steal the sensitive information. 
As we will discuss in \S\ref{sec:threats}, uninitialized 
memory in cache systems poses higher threats than 
uninitialized memory in general, because the former may always
propagate to the disk at the point of flush. Therefore, the memory safety for page cache is especially critical in file systems.     




\vspace{0.5em}

\noindent \textbf{\texttt{F-6:} Crash-consistency models for persistent storage 
often lead to uninitialized memory.} 
Many file systems use crash consistency models, 
such as journaling~\cite{ext3}, logging~\cite{log-structuring}, 
and shadow paging~\cite{WAFL}. For instance, JFS 
employs a synchronous writing strategy to log the 
storage operations and \texttt{inode}. This will ensure 
that the file system can always recover to a correct state 
at a crash. The support of crash-consistency
introduces new data structures and requires sophisticated 
synchronizations across different components, which involves high 
complexity and often incurs vulnerabilities. For example, 
JFS logs all relevant in-memory data structures but many of 
them may not be properly initialized. The uninitialized data 
may later be written to persistent storage, resulting in 
leakage of kernel information (\eg \texttt{CVE-2004-0181}).


\vspace{0.5em}

\noindent \textbf{\texttt{F-7:} The mismatch between VFS semantics and 
file system implementations introduces vulnerabilities.} {The correctness of VFS is crucial to the underlying file system, however, it attracts 
much less attention than the integrity checking of the actual file system implementations.}
Operating systems often incorporate a Virtual File System (VFS), 
which offers uniform interfaces for user 
applications and redirects operations from 
applications to the underlying file system. 
For things to work correctly, the underlying file
system must follow the specifications of the VFS.
This is, however, often violated and brings many 
vulnerabilities. 
{VFS vulnerabilities (such as \texttt{CVE-2015-1420 and CVE-2015-2925}) could affect a wide range of file system implementations. 
Many fault-tolerant file system designs, such as EnvyFS} \cite{envyfs}{, also rely upon the correct operations of the VFS layer. 
However, existing work mostly focused on the bug detection for each individual file system (e.g., ext4, btrfs)} \cite{kim:sosp2019, Xu-sp19, Pillai-osdi14}{, 
they rarely work on the correctness of the VFS layer} \cite{galloway2009model}{. Therefore, we believe applying bug detection techniques 
like model checking and fuzzing to VFS could also significantly improve file system security.}

\vspace{0.5em}

\noindent \textbf{\texttt{F-8}: {Functionalities for supporting use-context can become 
the attack vectors of mobile and networked file systems. }}
Unlike generic file systems, mobile file systems and networked
file systems need special functionalities to support their 
use contexts. Implementation of these functionalities 
often involves
vulnerabilities, which we detail below. 

Mobile files systems typically run on portable devices like 
smart phones, which can easily get lost and encounter physical 
attacks. To protect user data under physical attacks, many mobile 
file systems introduce encryption mechanisms. For instance, 
APFS leverages AES-XTS or AES-CBC to encrypt user files on 
demand~\cite{apfs-cryption}. F2FS, instead of directly providing 
file encryption, utilizes a cryptographic stacked file system 
(eCryptFS) to encrypt data. However, vulnerabilities can arise
in the implementation, often breaking the security of the 
encryption mechanisms. APFS leaks
encryption keys via the Disk Utility hints (\texttt{CVE-2017-7149})
and APFS allows reading of decrypted data via 
Direct Memory Access (\texttt{CVE-2017-13786}). The eCryptFS on Samsung 
KNOX 1.0 uses a weak 
key generation algorithm, which makes brute-force attacks feasible
(\texttt{CVE-2016-1919}). 

Networked file systems need communication modules based on 
protocols like HTTP and RPC to support remote data transfer 
and access authentication. Many vulnerabilities appear 
in the communication modules, due to mishandling of malformed or 
crafted network packets, such as zero payloads, rafted packet header,
and uninitialized packets. For example, GlusterFS allows RPC 
requests to create symbolic links (e.g., \texttt{gfs3\_symlink\_req}) 
pointing to file paths in the GlusterFS volume. This enables adversaries
to send crafted RPC to create arbitrary symbolic links in the storage server and then execute arbitrary code (\texttt{CVE-2018-10928}).

\vspace{0.5em}
\noindent
\fcolorbox{black}{Gray}{\parbox[c]{13.7cm}{
\textbf{Summary:} Core components in file systems, 
due to their high complexities, contribute a majority 
of the vulnerabilities. The situation is worsening because
new features are being integrated into the core components 
and are building up the complexities. 

\vspace{0.5em}

\textbf{Insight-2:} Different file system components introduce
vulnerabilities in semantic-dependent ways. This suggests that
different tools should be tailored, in a semantic-aware manner, 
to detect vulnerabilities in different components. For instance, 
to detect uninitialized memory in cache systems, a tool needs to 
identify not only explicit access (\eg by regular file system operations) 
but also implicit access (\eg by synchronization of cache management).

\vspace{0.25em}
\textbf{Insight-3:} At present, security seems insufficiently considered when 
new file system features are added. It is desirable that the trade-off 
between functionality/efficiency and security is taken into account 
before new features are incorporated, particularly in 
security-sensitive sectors. References about which new features
bring what security issues can be found from \textbf{\texttt{F-3}} 
to \textbf{\texttt{F-8}}. To fundamentally improve the situation, 
proactive measures, such as regression test~\cite{yoo2012regression, agrawal1993incremental,onoma1998regression} on the new features, 
must be taken.
}}
\vspace{0.5em}

\begin{figure*}[!t]

        \begin{subfigure}[b]{1\textwidth}
        \centering
        \includegraphics[width=0.6\linewidth]{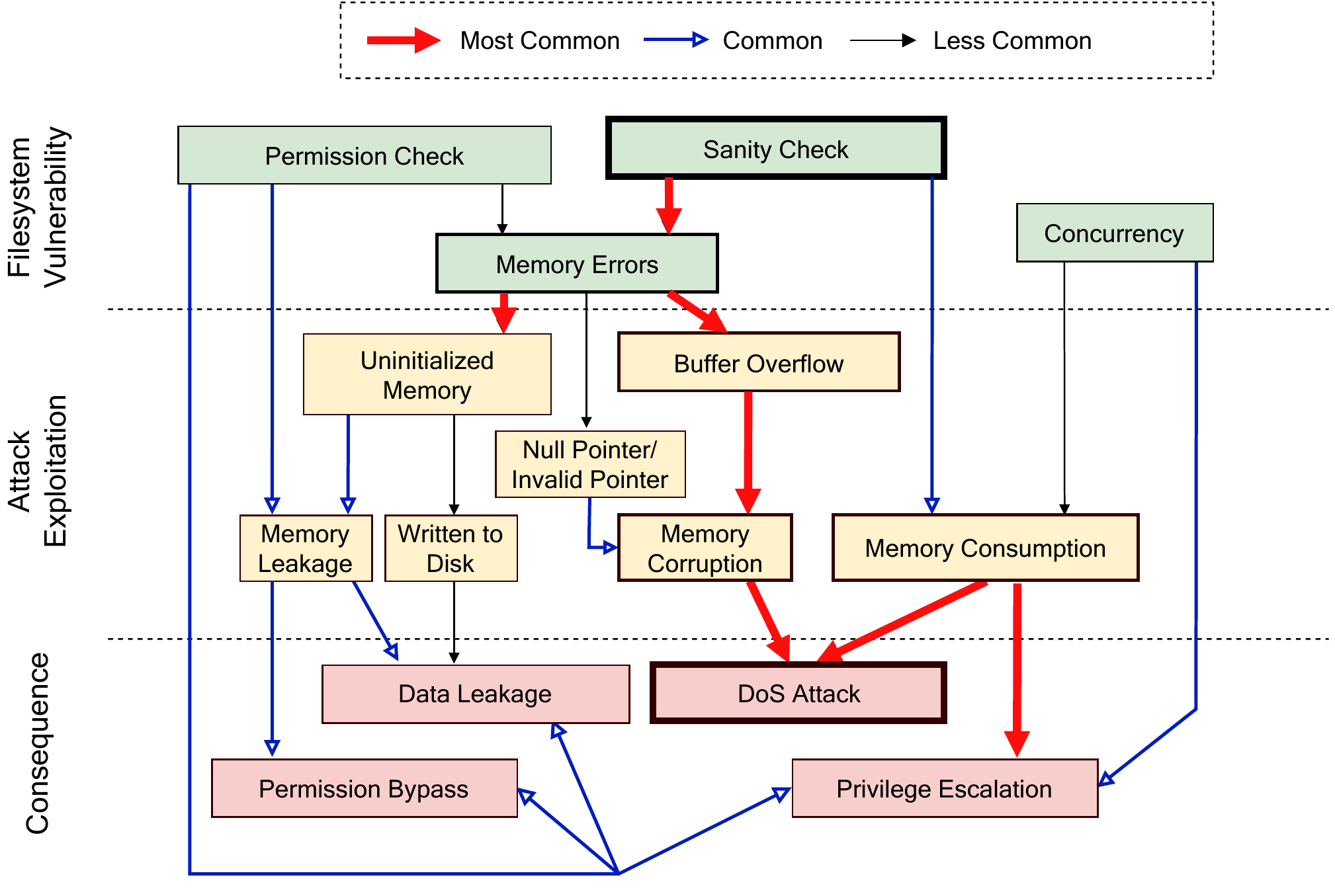}
        \caption{Threats to generic file systems}
        \label{fig:generic-threat}
        \end{subfigure}
        \begin{subfigure}[b]{1\textwidth}
        \centering
        \includegraphics[width=0.65\linewidth]{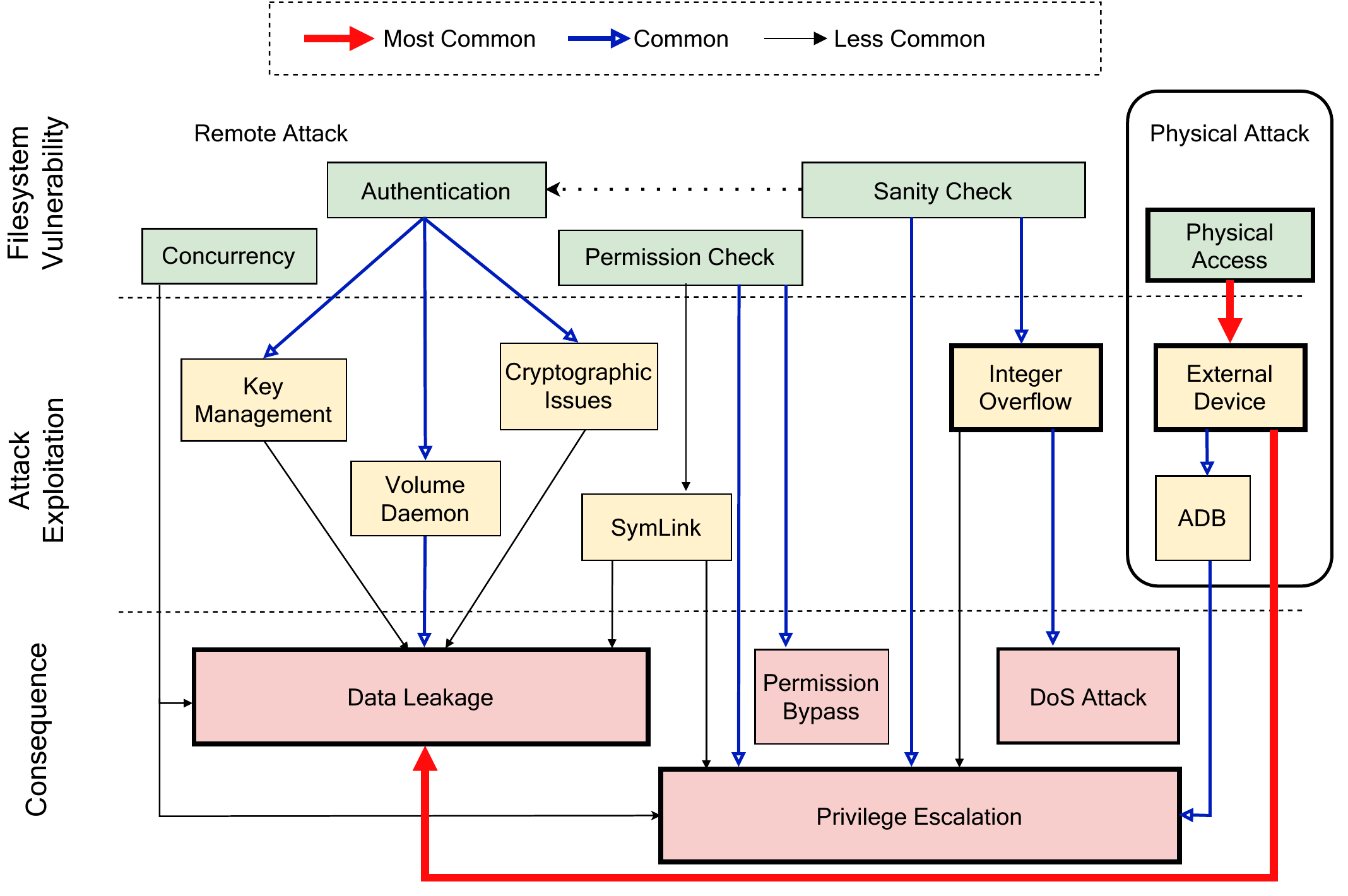}
        \caption{Threats to mobile file systems}
        \label{fig:mobile-threat}
        \end{subfigure}
        \begin{subfigure}[b]{1\textwidth}
        \centering
        \includegraphics[width=0.6\linewidth]{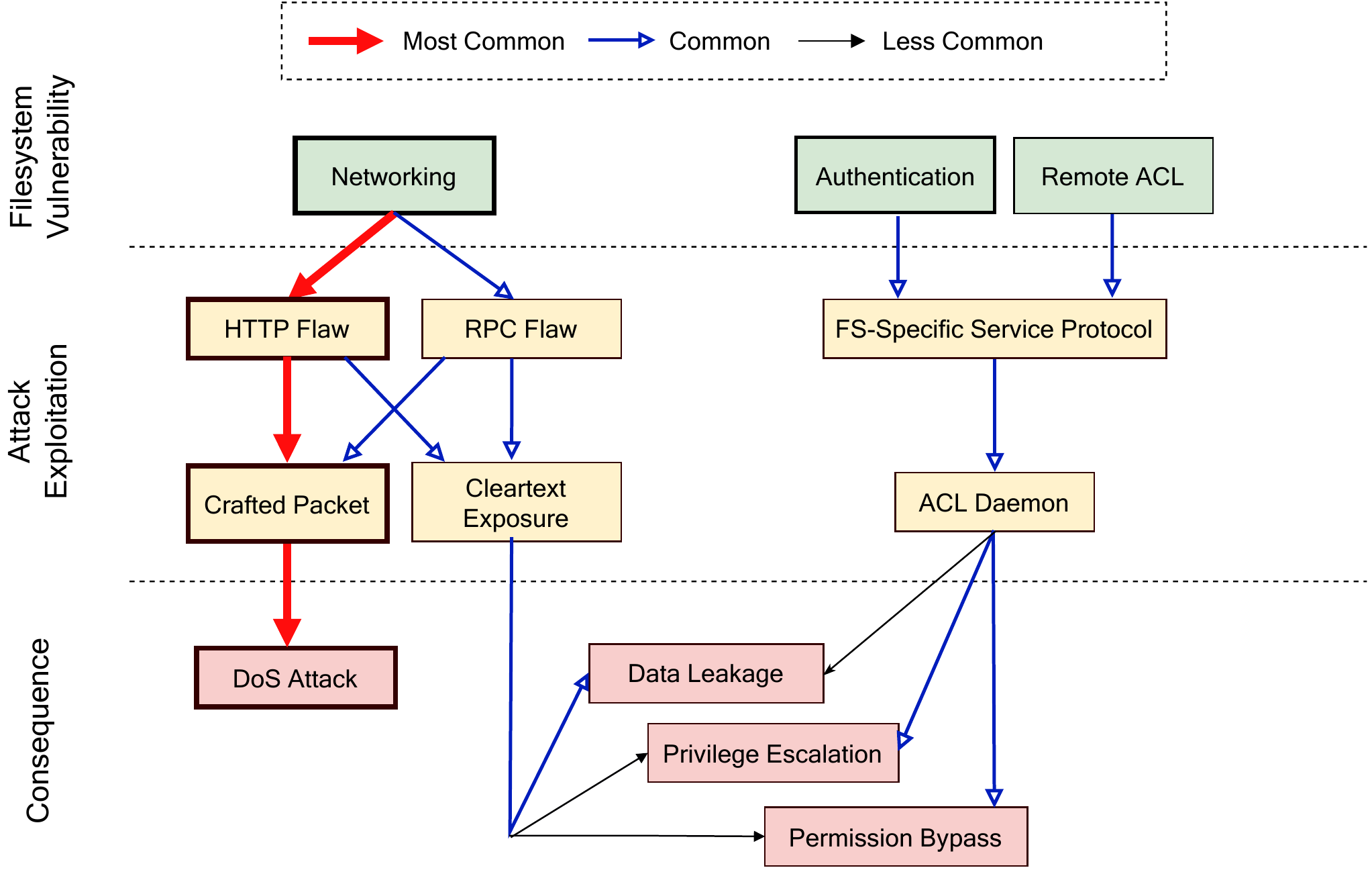}
        \caption{Threats to networked file systems}
        \label{fig:networked-threat}
        \end{subfigure}
        \caption{Trajectory from vulnerabilities to threats in different types of file systems.}
        \label{fig:fsattack}
\end{figure*}


\section{Threats of File System Vulnerabilities}
\label{sec:threats}

Our taxonomy brings insights towards mitigating vulnerabilities 
in file systems. However, file system vendors or distributors 
may still hesitate to take action because the threats of 
the vulnerabilities are unclear to them. In this section, we 
summarise the threats posed by file system vulnerabilities. 
We anticipate bringing evidence to motivate mitigation
and bring insights into prioritizing the mitigation against more
dangerous vulnerabilities. 

Overall, different vulnerabilities, when exploited, 
can lead to very diverse consequences. Figure~\ref{fig:fsattack}
gives an overview of the trajectories from vulnerabilities
in different file systems.
The major attacks of generic file systems are caused by incomplete sanity checks which 
exploit memory errors like memory corruption and memory contention, which often result in Denial-of-Service (DoS). 
Mobile file systems have their unique attack paths, which mainly exploit the physical access vulnerabilities. 
They could bring severe consequences, such as data leakage. 
As networked file systems launches multiple file system instances, due to their incomplete packet filtering and validation, 
their network connection could be attacked with crafted packets via HTTP or RPC protocols.  Table~\ref{tab:mobifs-summary}
presents the detailed distribution of vulnerabilities  
based on the threats they can pose. Following a similar organization
of \S\ref{sec:taxonomy}, we summarize our major findings below.

\begin{table}[t]
    \footnotesize
        \footnotesize
    \caption{Distribution of file system vulnerabilities based on their threats. {D:} DoS, {L:} Data Leakage, {E:} Privilege Escalation, {A:} Access Permission Bypass, {O:} Other (\eg write access to \code{proc} entries like \texttt{CVE-2004-2613}).}
    \label{tab:mobifs-summary}
    \centering
    \vspace{-2ex}
    \begin{tabular}{m{36pt}<{\centering}|m{52pt}<{\centering}|p{16pt}<{\centering}|p{16pt}<{\centering}|p{16pt}<{\centering}|p{16pt}<{\centering}|p{16pt}<{\centering}}
       \hline
        \textbf{Type} & \textbf{Name}  & D & L & E & A & O \\\hline

        \multirow{12}{*}{Generic}  & JFS~\cite{jfs} & \colorbox{pink}{50\%} & 25\%& 0\%& 25\% & 0\% \\
        
        &  Ext2~\cite{ext2} & \colorbox{pink}{57\%} & 14\%& 14\%& 14\% & 0\% \\
        
        & XFS~\cite{xfs} & \colorbox{pink}{70\%} & 22\% & 4\% & 4\% & 0\%  \\
        
        &  VFS~\cite{vfs} & \colorbox{pink}{50\%} & 4\% & 14\% & 25\% & 7\% \\
        
        &    procfs~\cite{procfs} &  \colorbox{pink}{44\%} & 22\% & 22\% & 6\% & 6\%\\
         & ReiserFS~\cite{reiserfs} & \colorbox{pink}{60\%} & 0\%& 40\%& 0\% & 0\% \\


        & Ext3~\cite{ext3} & \colorbox{pink}{78\%}& 11\%& 0\%& 11\% & 0\% \\
        
         & tmpfs~\cite{tmpfs} & 33\%& 17\%& \colorbox{pink}{50\%}& 0\% & 0\% \\


        & Ext4~\cite{ext4} & \colorbox{pink}{87\%} & 4\% & 7\% & 2\% & 0\% \\
        
        & Btrfs~\cite{btrfs} & \colorbox{pink}{54\%} & 15\% & 8\% & 23\% & 0\% \\

        & F2FS~\cite{f2fs} & \colorbox{pink}{94\%} & 0\% & 6\% & 0\% & 0\% \\ 
            
    \hline
 
    \multirow{5}{*}{Mobile}  &  HFS/HFS+~\cite{hfs} & 25\% & 13\% & \colorbox{pink}{62\%} & 0\% & 0\% \\
    
        & eCryptFS~\cite{ecryptfs} & 25\% & 25\% & \colorbox{pink}{38\%} & 0\% & 13\% \\

    & F2FS~\cite{f2fs}   & 18\% & \colorbox{pink}{55\%} & \colorbox{pink}{27\%} & 0\% & 0\%\\
    
    & APFS~\cite{apfs} & 0\% & 25\% & \colorbox{pink}{75\%} & 0\% & 0\%  \\

    & Others    &  \colorbox{pink}{28\%} & 17\% & \colorbox{pink}{33\%} & 11\% & 11\%\\
    \hline


   \multirow{6}{*}{Networked} & NFS~\cite{pawlowski2000nfs} & \colorbox{pink}{68\%} & 1\% & 12\% & 18\% & 0\% \\
    
    & GlusterFS~\cite{glusterfs} & \colorbox{pink}{28\%} & 17\% & 17\% & \colorbox{pink}{38\%} & 0\% \\
    
     & OpenAFS~\cite{openafs} & \colorbox{pink}{63\%} & \colorbox{pink}{30\%} & 4\% & 4\% & 0\% \\

    & HDFS~\cite{hdfs} & \colorbox{pink}{33\%} & 17\% & \colorbox{pink}{33\%} & 17\% & 0\% \\
    
    & CephFS~\cite{weil2006ceph} & \colorbox{pink}{73\%} & 20\% & 7\% & 0\% & 0\%\\
       \hline
    \end{tabular}
    \vspace{-3ex}
\end{table}

\vspace{0.5em}

\begin{figure}[t!]
        \centering
        \begin{minipage}[c]{\linewidth}
        \centering
                \includegraphics[width=0.85\linewidth]{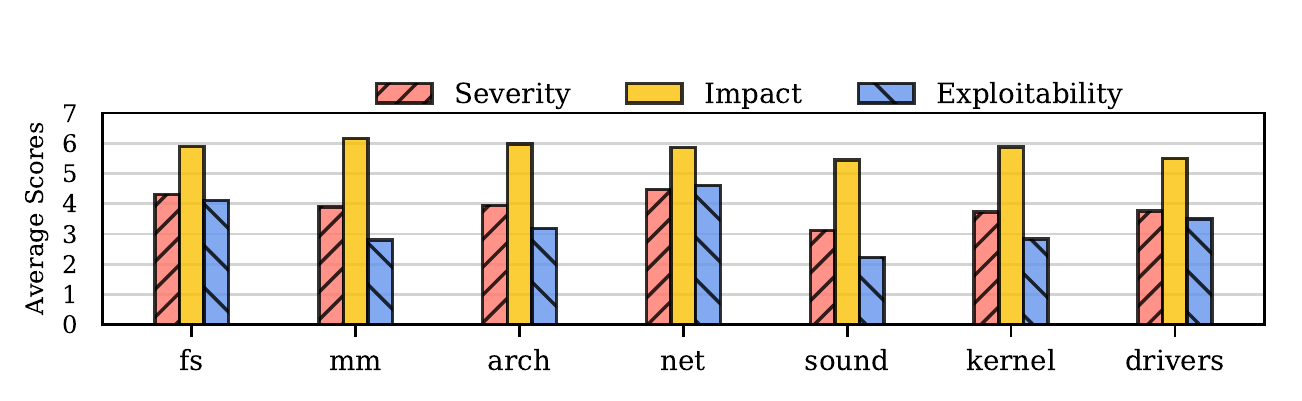}
        \end{minipage}
        \vspace{-3ex}
        \caption{Vulnerability comparison of core OS components. In the figure, \texttt{Severity}, \texttt{Impact}, and \texttt{Exploitability} are extracted from the CVSS Score, the CIA Impacts, and the Access Complexity in CVE reports, respectively.}
        \label{fig:kernel-scores}
        \vspace{-3ex}
\end{figure}

\noindent \textbf{\texttt{F-9:} File systems vulnerabilities 
impose a comparable level of security threat to vulnerabilities 
in other OS components.} Similar to our study 
on file system vulnerabilities, we gathered vulnerabilities 
in other operating system (OS) components that are committed to the CVE database 
in the past 20 years. As illustrated in Figure~\ref{fig:kernel-scores}, 
file system vulnerabilities are as dangerous as vulnerabilities in other 
components, considering severity, impact, and exploitability as 
the metrics. 


\vspace{0.5em}

\noindent \textbf{\texttt{F-10:}
DoS is the dominant threat brought by vulnerabilities to file systems.} 
About 61\% of all the vulnerabilities can lead to DoS. 
They often first incur one of the following behaviors
and eventually cause a DoS.

\begin{itemize}[leftmargin=*]

    \item \textit{Memory corruption}: 
    Memory errors and race conditions in file systems,
    as they typically do, can often cause memory corruption, 
    followed by a kernel panic, such as the case \texttt{CVE-2013-6382}. 
    Many semantic
    vulnerabilities can also lead to memory corruption
    and subsequent kernel panics. The previously discussed 
    \texttt{CVE-2015-8839} is such an example. 
    
    \item \textit{Memory consumption}: Another major 
    cause of DoS in file systems is excessive memory consumption. 
    Most of such cases are caused by memory leakage
    vulnerabilities. For instance, XFS before Linux 4.5.1  does not properly free the memory allocated for 
    extended file attributes  (\texttt{CVE-2016-9685}), which can be exploited to exhaust the memory and cause a DoS.
	
    \item \textit{System hang}: CPU over-consumption,
    mainly brought by infinite loops and deadlocks,
    can turn into DoS. For instance, the 
    integer overflow reported in \texttt{CVE-2017-18257} 
    can be exploited to trigger an infinite loop in the block allocation of F2FS, causing a system hang.
    
    \item \textit{Network congestion}: 
    Vulnerabilities derived from protocol flaws, 
    which account for about 16\% of all the vulnerabilities in 
    networked file systems, frequently 
    trigger DoS. As we pointed out before, such vulnerabilities 
    typically happen because of mishandling of malformed
    network packets, which often easily disrupt normal
    communication and result in DoS. For example, 
    the remote procedure call (RPC) module in 
    OpenAFS (before 1.6.23 or 1.8.x before 1.8.2)
    does not restrict the size of an input. 
    Adversaries can send, or claim to send, 
    large inputs and consume server resources 
    waiting for those inputs, blocking service to other 
    valid connections (\eg \texttt{CVE-2018-16949}).

\end{itemize}


\vspace{0.5em}

\noindent
\textbf{\texttt{F-11:} Data leakage is another big 
threat that vulnerabilities pose to file systems.} 
Three major types of file system vulnerabilities often 
lead to data leakage. The first type is uninitialized 
memory, where sensitive information, such as 
user data, file system metadata, and kernel memory
unintentionally propagates to unauthorized destinations
(\eg \texttt{CVE-2005-0400} and \texttt{CVE-2004-0177}).
\emph{In comparison to other software systems, uninitialized 
memory in file systems can more often lead to data 
leakage}. As we pointed out in \S\ref{sec:taxonomy}, 
file systems incorporate many cache systems to 
improve their performance. Memory objects used 
by the cache systems shall be flushed to the disk, 
which creates extra paths for uninitialized memory to leak.

The second type is permission-bypassing vulnerabilities. 
By exploiting such vulnerabilities, the adversaries
can gain access to files without the proper 
authorization and thus, 
indirectly leak private data. For instance, 
\texttt{CVE-2012-4508} allows adversaries to 
obtain sensitive information from a deleted file by 
exploiting a race condition to gain permissions. 

The last type is vulnerabilities in mobile file systems
that enable physical attacks. The most direct consequence 
of such attacks is the breach of private data.
For instance, physical adversaries, by exploiting 
\texttt{CVE-2014-7951} in Android Debug Bridge of Android 4.0.4, 
can gain a direct connection to the device to 
read/write arbitrary files in the file system.

\vspace{0.5em}

\noindent
\textbf{\texttt{F-12:} Attackers can bypass access control
via exploiting vulnerabilities in the enforcement of 
file permissions.} Flaws in the enforcement 
of permissions largely become vulnerabilities. In comparison 
to other types of vulnerabilities, permission-related
vulnerabilities can often be reliably exploited and
consistently lead to permission bypassing. 

Beyond the authentication of user identities, 
networked file systems also run access control 
on file access requests. Two categories of vulnerabilities 
in this type of access control can also be exploited
to gain file permissions on the removed servers: 
(1) validation operations before setting ACL are missing; 
(2) permission checks are missing. For example, the NFS client
in Linux 2.6.29.3 or before misses a check on the 
permission bits for execution, allowing local users 
to bypass permissions and execute files on an NFSv4 fileserver
(\texttt{CVE-2009-1630}). Similar issues are also reported 
in \texttt{CVE-2005-3623} and \texttt{CVE-2016-1237}.

\vspace{0.5em}

\noindent \textbf{\texttt{F-13:} Adversaries
can achieve privilege escalation via exploiting
file system vulnerabilities.} All the file systems
we studied run at a high privilege. A significant 
subset of sanity-check vulnerabilities can enable 
adversaries to improperly ``inherit'' the high privilege. 
Some representative examples are: (i) 
\texttt{CVE-2017-5551} preserves the \texttt{setgid}
bit during a \texttt{setxattr} call in a tmpfs file system, allowing 
local users to gain group privileges with restrictions on execute 
permissions; (ii) \texttt{CVE-2016-5393} wrongly gives HDFS service
privileges to users who can only authenticate with the HDFS NameNode. 
This enables arbitrary commands execution with the same privileges as the 
HDFS service; (iii) \texttt{CVE-2016-1572} does not validate mount 
destination file system types, which allows local users to gain privileges 
by mounting over a nonstandard file system.

Authentication protocol used in modern networked file systems 
can also fail due to vulnerabilities, and thus, give 
users undeserved privileges. Current networked file systems
either develops its own authentication protocol~\cite{weil2006ceph} 
or reuses the third-party infrastructure~\cite{steiner1988kerberos, pawlowski1994nfs}.
Both ways of authentication can be tricked and bypassed.  
We observe a large number of such cases in Ceph and HDFS, 
mostly caused by mishandling of user credential messages 
(\eg \texttt{CVE-2013-4134}, \texttt{CVE-2009-3516}).

\vspace{0.5em}

\noindent
\textbf{\texttt{F-14:} Vulnerabilities in mobile file systems 
can sabotage the protection of data privacy.} 
To secure private data, mobile file systems have adopted protection mechanisms, 
mainly including encrypting user files~\cite{apfs-cryption} and isolating critical data 
inside trusted execution environments (TEEs). However, 
both mechanisms can be bypassed due to vulnerabilities. 
Since vulnerabilities related to file encryption 
have been discussed in \S\ref{sec:taxonomy} 
(see \textbf{\texttt{F-8}}), we now focus on TEEs below. 

TEE has become a standard feature on mobile devices, 
which has been leveraged by mobile file systems 
to enable an isolated environment for protecting 
sensitive data. However, a drop-in deployment of 
TEE can be insufficient due to vulnerabilities in 
the interactions with TEE. For instance, a system crash
in the non-secure domain of a TrustZone-enabled 
ARM platform will trigger an interrupt to 
switch the system context to the secure domain. 
Therein, CPU, memory, and register states can be accessed. 
However, an armored adversary can prevent the non-secure 
domain to issue the context switch, and initiate a 
Man-in-the-middle attack to obtain the execution information. 
For another example,
eCryptFS uses an algorithm to generate a 32-byte AES key that combines 
the user password and 32-byte TIMA (TrustZone-based Integrity Measurement) key.
The vulnerability (\texttt{CVE-2016-1919}) occurs because 
\textit{Base64.getEncoder} expends the input with a ratio of 4:3. This means that only 24 bytes 
determine the final eCryptFS key 
and adversaries can easily crack the key with a brute-force search.

\vspace{0.5em}

\noindent
\textbf{\texttt{F-15:} Protocol flaws in networked file
systems can enable code injection attacks.} Consider CephFS as 
an example. It uses the Ceph Object Gateway daemon (RGW) as an 
HTTP server to communicate with Ceph storage cluster and 
retrieve file data. The HTTP response splitting 
(HRS) flaw (\eg \texttt{CVE-2015-5245}) in the RGW can enable 
the well-known Carriage Return Line Feed (CRLF) injection
and cross-site scripting (XSS) attacks. Once the malicious code 
is injected, attackers will have the privilege to compromise the 
victim server, and then tamper the entire networked storage cluster.

\vspace{0.5em}

\begin{figure}[t!]
    \begin{subfigure}[b]{1\textwidth}
        \centering
        \includegraphics[width=0.95\linewidth]{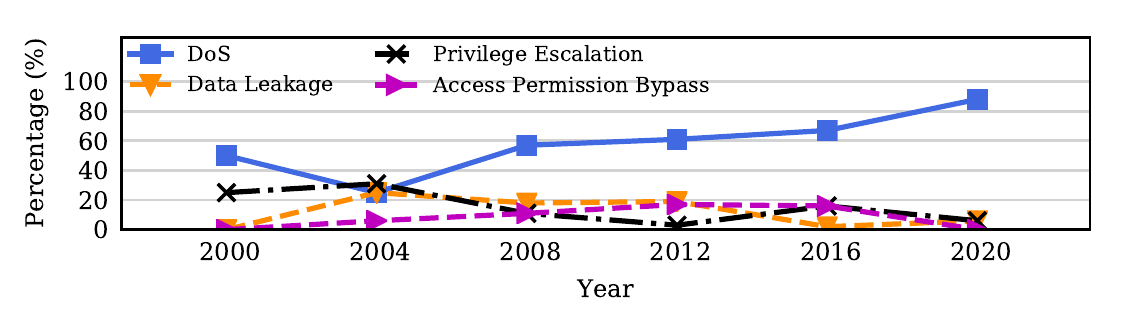}
        \label{subfig:gfsattacktrend}
        \vspace{-1ex}
        \caption{The threats of generic file system vulnerabilities.}
    \end{subfigure}
    \begin{subfigure}[b]{1\textwidth}
        \centering
        \includegraphics[width=0.95\linewidth]{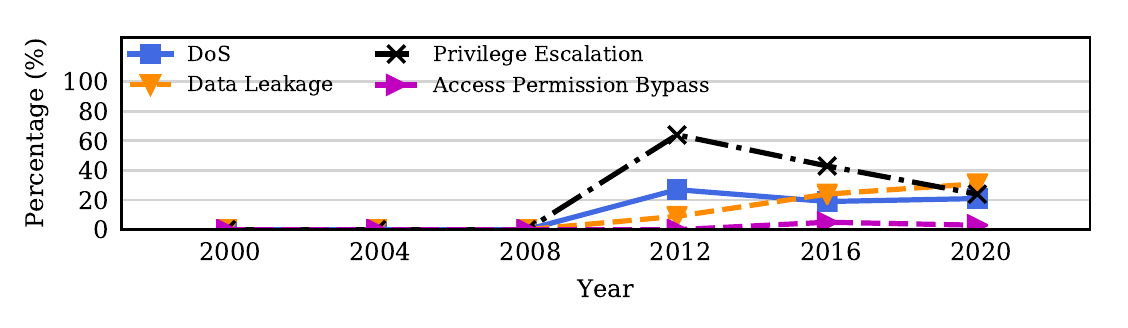}
        \label{subfig:mfsattacktrend}
        \vspace{-1ex}
        \caption{The threats of mobile file system vulnerabilities.}
    \end{subfigure}
    \begin{subfigure}[b]{1\textwidth}
        \centering
        \includegraphics[width=0.95\linewidth]{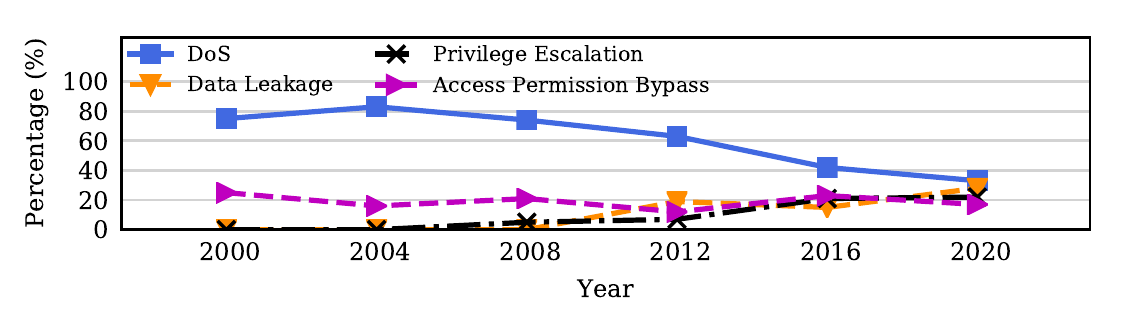}
        \label{subfig:dfsattacktrend}
        \vspace{-1ex}
        \caption{The threats of networked file system vulnerabilities.}
    \end{subfigure}
    \vspace{-8ex}
    \caption{Evolutionary trend of the threats imposed by file system vulnerabilities. 
    As for mobile file systems, no CVEs were reported before 2009.
    \vspace{-12ex}
    \label{fig:fsattacktrend}}
    
\end{figure}

\noindent
\textbf{\texttt{F-16:} Vulnerabilities in a single instance of networked file system 
can impose threats to the entire system.} 
\label{sec:correlation}
Networked file systems (\eg  HDFS~\cite{shvachko2010hadoop} and 
GFS~\cite{ghemawat2003google} ) can often run a group of parallel 
instances on a cluster of servers. Each instance runs atop 
the local file system and delegates the file system to store data 
on the disk. The consequence of vulnerabilities in a single instance 
can propagate to all other instances in the same group. 
Consider GlusterFS as an example. GlusterFS supports \emph{brick multiplexing} 
to reduce memory consumption, which allows multiple compatible bricks to 
share the same process and manage their own data volumes. However, 
a \emph{NULL} pointer dereference (\texttt{CVE-2018-10914}) triggered 
by a client request in one instance can interrupt the brick process. 
This will propagate to other multiplexed brick processes, crushing 
the entire GlusterFS. Other similar cases include \texttt{CVE-2012-4417}, \texttt{CVE-2019-15538}, \texttt{CVE-2020-24394}.

\vspace{0.5em}

\noindent
\textbf{\texttt{F-17:} Vulnerabilities in different types 
of file systems present a different evolutionary trend in their
threats.} As shown in Figure~\ref{fig:fsattacktrend}, vulnerabilities 
in \textit{generic} file systems tend to result in more DoS and fewer 
attacks of other types. However, vulnerabilities in \textit{mobile} 
file system and \textit{networked} file systems present an opposite 
evolutionary trend. A presumable reason is vulnerabilities in generic 
file systems are more conventional and their consequences are limited by 
mitigation mechanisms such as memory sanitizer \cite{serebryany2012addresssanitizer} and thread sanitizer \cite{serebryany2009threadsanitizer}. In contrast, 
many vulnerabilities in mobile/networked file systems belong to 
newly emerging types (\eg protocol flaws) and have been less mitigated.
Considering all types of file systems together, the evolution in the 
threat of their vulnerabilities presents no clear trend. 

\vspace{0.5em}
\noindent
\fcolorbox{black}{Gray}{\parbox[c]{13.7cm}{
\textbf{Summary:} 
The threats of file system vulnerabilities, varying across different 
types, ranging from more functionality-critical
ones like DoS to more security-critical ones like privilege escalation. 
History data shows that file system vulnerabilities, in general, 
are not becoming less harmful. 

\textbf{Insight-4:} Our study unveils that different file system 
operations, when becoming vulnerable, can turn into different threats. 
This brings insights towards the mitigating process (\eg patching):
vulnerabilities involved in operations that can result in higher 
threats shall be prioritized. For instance, uninitialized memory 
errors in cache systems, which often incur data leakage, 
deserve an earlier fix than excessive memory consumption 
that typically incurs DoS. References about which file 
system operations can lead to what threats can be found from 
\textbf{\texttt{F-10}} to \textbf{\texttt{F-16}}.

\textbf{Insight-5:} Our study also unveils that vulnerabilities 
in mobile/networked file systems are presenting increasing 
threats (\eg more data leakage than DoS). Considering the 
increasing popularity of  mobile/networked file systems in the 
coming era of IoT, our study brings evidence that more efforts 
should be prioritized to mitigate vulnerabilities in 
mobile/networked file systems.
}}

\section{Patching of File System Vulnerabilities}
\label{sec:patching}

Our taxonomy and threat analysis offer insights and motivations
to mitigate file system vulnerabilities. We take a step further to analyze their patches and gain deeper insights. This will be beneficial to file system developers,
since it can help people learn about the common patterns of patching file
systems and guide them to avoid the deficiencies in today's practice.
In the patch analysis process, we aim to answer three questions: 
(1) \emph{how complex are the patches?} (2) \emph{how are the patches applied,
systematically or in an ad-hoc manner?} (3) \emph{are patches across different file
systems correlated, and how that may affect the patching process?}

\vspace{0.5em}

\begin{figure}[t]
        \centering
        \includegraphics[width=.8\linewidth]{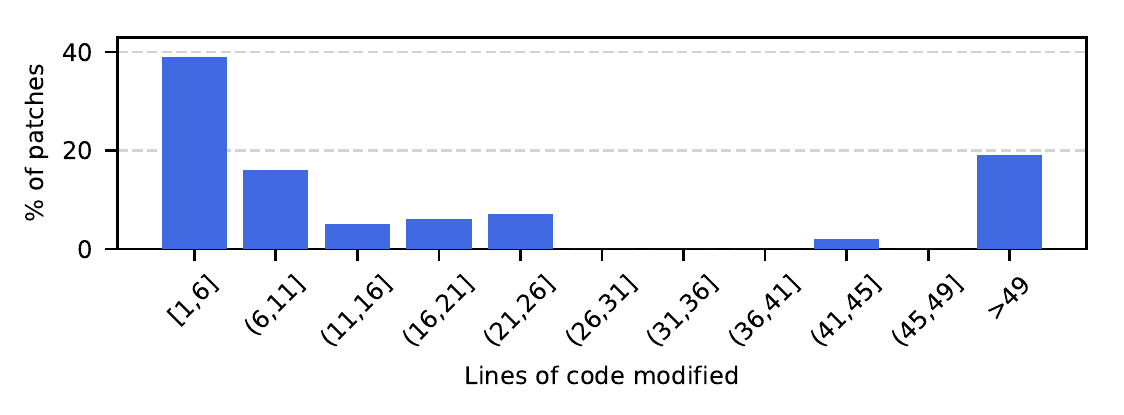}
        \vspace{-3ex}
        \caption{Distribution of file system patches, based on the lines of code (LoC) changed.}
        \label{fig:patch_bar}
\end{figure}

\noindent
\textbf{\texttt{F-18:} Most of the patches ($56\%$) produce small modifications to the file systems.} 
Figure~\mbox{\ref{fig:patch_bar}} shows the distribution of patches based on their complexities (i.e., how many lines of code are modified). 
We find that a large number of patches are simply adding a check to the memory boundary
(\eg \texttt{CVE-2013-6382}) or fixing a syntax error 
(\eg \texttt{CVE-2013-1848}). These patches constitute the low-complexity group. 
Most of the strategies are conventional, such as adding boundary 
checks to fix buffer overflows (e.g., \texttt{CVE-2008-3531}). 
For instance, the patches of many race conditions simply make the responsible
operations atomic. The drawback of such ad-hoc strategies is
they are too passive. They can neither help prevent similar vulnerabilities in the codebase nor bring global benefits like new designs. We believe a more systematic strategy should be taken (see our discussion in \textbf{F-20}).




\vspace{-2ex}

\vspace{1em}

\noindent
\textbf{\texttt{F-19:} A non-negligible portion ($20\%$) of the patches involve complex modifications, such as fixes to multiple file systems and comprehensive validations of file system data structures. Some vulnerabilities that are shared across file systems can cause duplicated patches.}
In most of the patches we studied, the patches aim to solve a unique 
flaw in a certain file system. In the study, we discovered that for a 
shared vulnerability in multiple file systems, developers release patches 
that have a similar strategy or even the same code logic. 
These patches build up the high-complexity group.
Strategies of 
a certain patch to a specific file system can be borrowed by patches 
that are used by other types of file systems.
For example, as shown in Figure~\mbox{\ref{fig:fs-patch}}a, the VFS subsystem added a new function \texttt{setattr\_killpriv} to perform additional permission and sanity checks for the extended attributes (xattrs). 
Since most Linux file systems supports xattrs, they have to properly invoke this function in their own $setattr()$ to ensure security. As a result, these file systems generate similar patches (Figure~\mbox{\ref{fig:fs-patch}}b and c, \texttt{CVE-2015-1350}).
Such pattern also occurs in the patching of \texttt{xattr} block caching in the ext2 and ext4 
file systems. Both of the file systems have to convert the original meta block cache from \texttt{mbcache} to \texttt{mbcache2}. Hence, they should add the same sanity checking to the entries and generate similar patches (\eg \texttt{CVE-2015-8952}).

A cross-filesystem patching approach could offer two major benefits.  First, having a unified solution 
simplifies the patch development process as well as the future 
maintenance. Second, file systems which share the patch can 
take advantage of further improvements in each individual file system, 
which enables us to improve the overall robustness and reliability.


\begin{figure*}[!t]
        \begin{subfigure}[b]{0.98\textwidth}
        \centering
        \begin{lstlisting}[language=C,style=base]
        + int setattr_killpriv(struct dentry *dentry, struct iattr *iattr)
        \end{lstlisting}
        \caption{A new interface added to VFS that removes extended privilege attributes.}
        \label{fig:vfs-update}
        \end{subfigure}
        \par\bigskip
        \begin{subfigure}[b]{0.49\textwidth}
        \centering
        \begin{lstlisting}[language=C,style=base]
int ext3_setattr(struct dentry *dentry, struct iattr *attr)
  if (error)
    return error;
@+ error = setattr_killpriv(dentry, attr);
+ if (error) 
+       return error; @
        \end{lstlisting}
        \caption{Patch to ext3 filesystem.}
        \label{fig:ext3-patch}
        \end{subfigure}%
        \hfill
        \begin{subfigure}[b]{0.49\textwidth}
        \centering
        \begin{lstlisting}[language=C,style=base]
int btrfs_setattr(struct dentry *dentry, struct iattr *attr)
  if (err)
    return err;
@+ err = setattr_killpriv(dentry, attr); 
+ if (err)
+   return err; @
        \end{lstlisting}
        \caption{Patch to btrfs filesystem.}
        \label{fig:btrfs-patch}
        \end{subfigure}
        
        \caption{Similar patches to different Linux file systems to reflect the same VFS inference update.}
        \label{fig:fs-patch}
\end{figure*}

\noindent
\vspace{0.5em}

\noindent
\textbf{\texttt{F-20:} Some patches demonstrate systematic strategies, which can benefit vulnerability prevention.}
Despite ad-hoc strategies still being dominant, we do observe 
systematic strategies in two cases. First, there are patches that perform systematic verification on file system data structures instead of simply applying an ad-hoc fix. For instance, \texttt{CVE-2018-14613} reports a simple invalid pointer dereference issue when mounting and operating a crafted \code{btrfs} image caused by invalid block group items. This specific vulnerability could have been solved by a low complexity patch. However, the patch to it demonstrates good practice by adding a new sanity check function that performs a complete check of the item size, offset, object id, type, and used space. This not only helps fix the reported vulnerabilities, 
but also removes the undiscovered hidden flaws to prevent future vulnerabilities. Indeed, none of the follow-up CVEs has related issues when mounting and operating on \code{btrfs} images. 
Second, many patches of race conditions switch from lock-based 
implementations to lock-free implementations, such as deploying semaphore-based solutions (\eg \texttt{CVE-2014-9710}) or avoiding race conditions in the first place (\eg \texttt{CVE-2014-8086}). These patches bring design-wise insights, which are 
applicable to a more fundamental resolution of race conditions.

\vspace{0.5em}

\noindent
\textbf{\texttt{F-21:} Patches that modify the VFS need coordinated updates to the underlying file systems.} A patch to the VFS subsystem often requires updating the underlying file systems. Otherwise, the patch may not come into effect. In \texttt{CVE-2015-1350}, the VFS subsystem provides an incomplete set of requirements for \texttt{setattr} operations , which under-specifies the removal of extended privilege attributes. The corresponding patch silently introduces a new VFS API to fix the flaw. Concurrently, the patch has to modify all 21 affected underlying file systems to use the new API. While, in this example, the complexity incurred by the need for coordinated patching is well handled, it is not always so. For example, the patch to \texttt{CVE-2016-7097} modifies VFS but fails to update the complete set of 15 affected underlying file systems, leaving the vulnerability still open (reported as \texttt{CVE-2017-5551}). Until one year later, another patch to \texttt{CVE-2017-5551} eventually fixed the problem. 



\vspace{0.5em}
\noindent
\fcolorbox{black}{Gray}{\parbox[c]{13.7cm}{
\textbf{Summary:} Patches to file system vulnerabilities carry different complexities. 
The patches also adopt various strategies. Most of them 
are ad-hoc, bringing no help with addressing similar but yet unknown 
vulnerabilities. However, patches following systematic strategies
are arising. These patches present preventive measures or new designs, 
offering insights towards fundamentally mitigating vulnerabilities of 
the same root causes. Finally, file systems are often correlated. 
However, today's practice often overlooks the correlated file systems
when patching the target ones, thus leading to insecure patches.

\textbf{Insight-6:} Our study unveils that the current practice of patching
file systems is largely passive. Continuing this practice will likely prolong
the arm race between vulnerability creation and patching. To fundamentally 
escalate the security of file systems, the community should encourage and 
promote the adoption of preventive and secure-by-design patching. 
Exemplary strategies observed by our study have been presented in \textbf{\texttt{F-20}}.

\textbf{Insight-7:} Better attention needs to be drawn on the correlation 
of file systems when developing patches. To ensure the correctness of patches, 
any modification to a file system (in particular VFS) must be validated
to understand its impacts on other file systems. Whenever such impacts arise, the affected file systems must be updated accordingly. 
}}
\vspace{0.5em}

\section{Suggestions}
\label{sec:suggestions}

\begin{table}[!t]
    \footnotesize
        \caption{Suggested strategies to patch common types of file system vulnerabilities. The third column shows the percentage of CVEs that can be fixed with the suggested strategies.}
        \vspace{-2ex}
    \label{tab:linux-fix}
    \centering
    \begin{tabular}{p{30pt}<{\centering}|p{170pt}<{\raggedright}|p{80pt}<{\raggedright}}
                \hline
                \textbf{Type} & \textbf{Suggested Defense Strategies$^{1}$} & \textbf{\% of Applicable CVEs} \\
        \hline
            \multirowcell{5}{Sanity \\ checks}
                                             & FS image validation before mounting * &  48.2\%\\
					     & Validating fs parameters upon file access & 12.5\% \\
                                             & Memory pointer validation and bounds checking & 5.4\%\\
                                             & Handle exceptions of division by zero & 3.9\% \\
                                             & Avoid integer overflows & 3.6\% \\
        \hline
            \multirowcell{2}{Conc. \\ issues}
                                            & Enforce atomic ToCToU race operations &  42.9\%\\
                                            & Support of lock-free optimizations * & 28.6\% \\
        \hline
            \multirowcell{3}{Memory \\errors}
                                            & Initialize unused, allocated memory & 38.5\% \\
                                            & Initialize data before copying them from user space & 24.2\% \\
                                            & Reinitialize fs metadata upon file updates & 14.3\% \\
        \hline
            \multirowcell{3}{File \\ pems.}
                                            & Check enforcement of user privileges  & 42.9\% \\
                                            & Check file permissions for direct storage access  & 28.6\% \\
                                          & Validate fs metadata information before mounting fs  & 4.8\% \\
        \hline
        \multicolumn{3}{ l }{$^{1}$~* indicates the mitigation is systematic.}
    \end{tabular}
\vspace{-3ex}
\end{table}

\begin{table}[t]
    \footnotesize
	\caption{Suggested defense strategies for mobile \textit{fs} vulnerabilities. The second column shows the percentage of CVEs that could be fixed with the suggested strategies.  
	\vspace{-2ex}
	}
    \label{tab:mobfixes}
    \centering
    \begin{tabular}{|r|p{80pt}<{\raggedright}|}
                \hline
	    \textbf{Mobile Defense Strategies} & \textbf{\% of Applicable CVEs}\\
        \hline
        Enhance authentication mechanisms in ADB channels & 25.0\% \\
        Enhance USB debugging mode & 18.8\% \\ 
	    Enforce secure interaction between TrustZone and FS & 12.5\% \\
        \hline
	    (Total) & 56.3\% \\
	    \hline
    \end{tabular}
\vspace{-3ex}
\end{table}

\begin{table}[t]
    \footnotesize
        \caption{Suggested defense strategies for networked \textit{fs} vulnerabilities. The second column shows the percentage of CVEs that could be fixed with the suggested strategies.   
        }
        \vspace{-2ex}
    \label{tab:distfixes}
    \centering
    \begin{tabular}{|m{175pt}<{\raggedleft}|p{80
pt}<{\raggedright}|}
                \hline
                \textbf{Unique Defense Strategies} & \textbf{\% of  Applicable CVEs.}  \\
        \hline
        Enforce strict authentication/permission control & 26.4\% \\
	Verify (HTTP/RPC) network packets  & 23.1\%\\
	    Enforce VFS specifications  & 9.3\% \\    
	Fault isolation between \textit{fs} instances & 7.1\% \\ 
        \hline
	    (Total) & 65.9\% \\
	    \hline
    \end{tabular}
\end{table}

With this study, we discover a group of vulnerabilities 
that \textbf{our community is not handling well at this point}.
Most of them are semantic related or use-context related.
We summarize the common suggested strategies used by 
the patches on generic file system vulnerabilities in Table~\ref{tab:linux-fix}. 
The common defense strategies discussed of patching generic file system can be also applied to fix $43.7\%$ of the mobile file system and $34.4\%$ of the networked file system vulnerabilities. Meanwhile, we identify their unique defense strategies vulnerabilities, as shown in Table~\mbox{\ref{tab:mobfixes} and \ref{tab:distfixes}}. 

\vspace{0.5em}
\noindent
\textbf{\texttt{S-1:} Verifying file system implementations with semantics specified in VFS.}  
As discussed, VFS provides a uniform abstraction level for upper-level programs to interact with 
real file system implementations. It offers semantics (e.g., POSIX) for accessing the underlying file systems. 
However, the real filesystem implementation may not exactly satisfy the semantics, due to 
the unclear specifications between the VFS and low-level filesystem implementations. 
Given that VFS specifications have been defined, they can be refined to 
verify the filesystem implementations. 
Prior verification studies of file systems~\cite{nelson:sosp2017, sigurbjarnarson-osdi16} mostly worked on 
the crash safety. Our findings reveal that more efforts are required to achieve the end-to-end verification 
of each semantic mapping from the high-level VFS specification to the low-level filesystem implementation. 

\vspace{0.5em}
\noindent
\textbf{\texttt{S-2:} Enhancing secure channels between mobiles and external physical devices.} 
To defend against physical attacks, we need to continue developing secure channels between mobile devices and external devices such as Thunderbolt adapters and USB ports. According to our study, 
nearly 23\% of vulnerabilities in Android can be exploited by evil-maid attacks. These attacks can be prevented by 
explicitly authorizing the mounting of external devices. 

\vspace{0.5em}
\noindent
\textbf{\texttt{S-3:} Enforcing secure communication between TEE and file systems.} 
We still cannot solely rely on TEE or encrypted file systems or even both to 
ensure the security of mobile file systems. Particularly, we have to ensure the secure interactions between 
TEE and file systems, such as preventing the direct access to the memory used by Volume Daemon or storage 
management process from TrustZone. We have to carefully refine the interfaces between them, 
which can be improved with the development at both TEE and file system side~\cite{sok-oakland2020, android-ecosystem, android:csur}.

\vspace{0.5em}
\noindent
\textbf{\texttt{S-4:} Verifying network packet.} 
According to our study, 
verifying the network
packets, especially those via the HTTP or RPC protocols, can significantly reduce the risk
of network attacks in networked file systems.
The semantic specifications from upper-level service protocols defined by network file systems  
will facilitate such network packet verification. Similar network verification techniques
have been developed in software-defined networking domain~\cite{plankton:nsdi2020, networkv:popl2016}
and packet level authentication~\cite{p4v:sigcomm2018}, however, few of them focused on the 
semantic-aware verification of storage traffic from networked file systems. 

\vspace{0.5em}
\noindent
\textbf{\texttt{S-5:} Fault isolation for networked filesystem instances.} 
As discussed in $\S$\ref{sec:correlation}, a filesystem instance (e.g., a data volume manager) has 
correlations with many system components such as local file systems and OS kernel. These correlations 
significantly complicate the management of filesystem instances and increase the attack surface. 
Providing a fault isolation mechanism for filesystem instances can significantly enhance the 
security of the entire networked system. Systems techniques, such as containers~\cite{container:acsac18, xcontainer:asplos2019}, 
sandbox~\cite{sandbox:ccs2010}, and TEE~\cite{libseal:eurosys18}, have been proposed to enforce the isolation 
between programs. They can be leveraged to enable isolation between filesystem instances. 
However, a holistic approach that requires fine-grained function partition and placement for networked file systems 
is still highly desirable~\cite{correlated:osdi14}.

\section{Related Work} 
\label{sec:related}

\noindent{\bf Study of Bugs and Vulnerabilities.}
Past research has conducted many studies on defects in 
operating systems. The studies vary in targets. 
Many of them~\cite{lu-fast13, min-sosp15, Xu-sp19, wei2005tocttou} 
focus on functionality bugs in file systems while some others~\cite{huang-atc16, 
govindavajhala2003using} concentrate on virtual memory management problems. 
The studies also vary in scope that ranges from concurrent issues~\cite{fonseca-dsn10, 
lu-asplos08, leesatapornwongsa-asplos16, Gunawi-socc14} to specific families of 
vulnerabilities~\cite{chen-apsys11, lu2017unleashing,schumilo2017kafl}. 
Our work uses a methodology similar to many of the studies. 
However, we focus on the secuirty vulnerabilities in file systems. 
Unlike the previous file system studies that focused on bug causes and consequences~\cite{anonymous, lu-fast13}, 
our study centers around the understanding of the attack surface of file systems and 
the vulnerability exploitation procedure, while covering a wider range of CVEs across different types of file systems. 
To the best of our knowledge, our study is the first in-depth work of its kind.

\noindent{\bf Bug and Vulnerability Detection.} 
Research in this line can be classified into two categories:
bug finding and formal verification. Bug finding tools, such 
as FiSC~\cite{yang-osdi04}, \textsc{eXplode}~\cite{yang-osdi06}, 
and \textsc{juxta}~\cite{min-sosp15}, can discover file system 
bugs based on semantic-aware patterns. Our study can benefit
these tools by providing insights into expanding their patterns.
Formal verification~\cite{nelson:sosp2017, gu-osdi16,amani-asplos16,
sigurbjarnarson-osdi16,bornholt-asplos16,chen-sosp15,chen-sosp17}
is also promising. However, at this stage, it is still challenging to 
verify the security of an entire file system due to the high complexity~\cite{sigurbjarnarson-osdi16, xu-cav16, gu-osdi16}. 
Our study can complement formal verification. It pinpoints 
the file system components that are more vulnerable, thus
enabling formal verification to narrow down the scope 
and scale up. 
  
\noindent{\bf Secure File Systems.}
Besides bug study and detection, researchers have also been 
endeavoring to build secure and reliable file systems~\cite{lu-osdi14,Min-atc15}. 
Lu \etal~\cite{lu-osdi14} proposed the physical disentanglement to 
minimize storage faults propagation. Min \etal~\cite{Min-atc15} 
leveraged transaction flash storage to build crash-consistent file systems. 
Our study can bring insights to follow-up works in this line of research. 
For instance, we find that many vulnerabilities in networked file systems  
are caused by the lack of fault isolation across different instances.
This can motivate efforts to develop fault isolation mechanisms for file systems. 


\section{Conclusion}
\label{sec:conclusion}

This paper presents an empirical study on the security 
of modern file systems, following the angle of inspecting
vulnerabilities disclosed from mainstream file systems. 
Throughout the study, we build the first systematic understanding
of the attack surfaces faced by file systems, the threats
tied to the attack surfaces, and the limitations in today's 
practice of addressing the attack surfaces. We envision 
that our study will raise awareness of file system security
and, more importantly, offer insights towards improving
file system security.




\newpage
\balance
\small
\bibliographystyle{plain}
\bibliography{ref}

\end{document}